\documentclass[aps,pra,twocolumn,superscriptaddress,notitlepage,floatfix]{revtex4-2}
\usepackage{graphicx}
\usepackage{hyperref}
\usepackage{amsmath}
\usepackage{amssymb}
\usepackage[all]{hypcap}
\usepackage{CJKutf8}

\hypersetup{
	colorlinks=true,       
	linkcolor=blue,        
	citecolor=blue,        
	urlcolor=blue,        
	pdfborder={0 0 0}      
}

\begin{document}
	
	\title{Nonlinear dynamics in an artificial feedback spin maser II: dependence on feedback configurations and time crystal effects}
	
	\author{Lan Wu}
	\author{Qingchang Li}
	\author{Weiyu Wang}
	\affiliation{National Time Service Center, Chinese Academy of Sciences, Xi’an 710600, China}
	\affiliation{University of Chinese Academy of Sciences, Beijing 100049, China}
	
	\author{Qianjin Ma}
	\affiliation{National Time Service Center, Chinese Academy of Sciences, Xi’an 710600, China}
	
	\author{Guobin Liu}
	\affiliation{National Time Service Center, Chinese Academy of Sciences, Xi’an 710600, China}
	\affiliation{University of Chinese Academy of Sciences, Beijing 100049, China}
	\affiliation{Key Laboratory of Time Reference and Applications, Chinese Academy of Sciences, Xi’an 710600, China}
        \email{liuguobin@ntsc.ac.cn}
	
	\date{\today}
	
	\begin{abstract}
Artificial feedback spin masers exhibit rich nonlinear dynamics, yet a systematic framework for understanding and controlling their behaviors remains elusive. In this paper, we consider the polarity and geometric configuration of the feedback field as important control parameters for studying the nonlinear dynamics of spin masers. By cooperatively tuning the strength and phase delay of feedback field, we investigate the spin dynamics at various configurations of the feedback field. Periodic, quasiperiodic and chaotic spin oscillations are uncovered. Specifically, a unique spin oscillation mode akin to the time crystal effects emerges within a certain parameter regime. This work presents a different viewpoint understanding the nonlinear dynamics of spin masers including the time crystal.
	\end{abstract}
	
	\maketitle
	
\section{Introduction}

Spin masers or Zeeman masers are based on the optically pumped atomic magnetometers and were first realized in the $^3$He nuclear spin system \cite{Dehmelt1957, Bloom1962,Robinson1964, Richards1988}. Spin-exchange optical pumping provides an efficient way to prepare highly polarized noble-gas nuclear spins with the help of alkali metals \cite{Walker1997}, leading to the development of high-performance spin-exchange optical pumping spin masers \cite{Chupp1994, Stoner1996, Bear1998,Yoshimi2012,Sato2018}. Spin masers are widely used in fundamental physics and practical applications ranging from the tests of atomic electric dipole moments, Lorentz symmetry to non-destructive detection of metal material defects \cite{Rosenberry2001, Bear2000,Jiang2021,Bevington2019,Bevington2020}.

The dynamical behavior of the spin maser is typically described by the Bloch equations with an additional radiation damping or feedback driving term and shows complicated nonlinear dynamics such as chaos under strong radiation damping or feedback magnetic fields \cite{Lin2000,Abergel2002}. The introduction of artificial feedback mechanisms offers a flexible way of tuning the driving fields \cite{Yoshimi2002}. Previously, we have revealed the rich nonlinear spin dynamics by introducing a time delay feedback model in such an artificial feedback spin maser system \cite{Feng2025}. While most of the nonlinear effects predicted in the time delay model have been observed in experiments, the recently reported observation of time crystal phenomenon in spin masers can not be explained by the present theory \cite{Wang2025}. Recent reports also indicate that the inhomogeneity of the static magnetic field plays a key role in the formation of time crystal effects in spin masers \cite{Wang2024CP,Huang2025NC}. 

To address this issue, we try to understand the nonlinear dynamics of spin maser from a different point of view. In this paper, we consider two other key factors: the polarity of feedback field and the geometry of detection-feedback configuration. For the polarity, there are two situations: positive feedback and negative feedback. For the geometry of detection-feedback configuration, there are also two typical situations: 1) detect the $x$ component of the transverse spin magnetization and apply the feedback field in the $x$ direction, and 2) detect the $x$ component of the transverse spin magnetization and apply the feedback field in the $y$ direction. Using the time delay model, we show that various nonlinear effects including time crystal can be produced in the artificial feedback spin maser. We show that the polarity and geometry configurations of the feedback fields are probably the main reasons for the time crystal effects in the artificial feedback spin masers.
 
\section{MODEL}

We consider an ideal ensemble of single-species atomic spins subject to a static magnetic field and a delayed feedback field in the transverse plane. The transverse feedback field is described by the two-dimensional vector $\mathbf{B}^{\mathrm{fb}} = (B_x^{\mathrm{fb}}, B_y^{\mathrm{fb}})$. The total magnetic field is $\mathbf{B} = (B_x^{\mathrm{fb}}, B_y^{\mathrm{fb}}, B_0)$, with $B_0$ the magnitude of the static field along the $z$ axis. The spin magnetization vector is $\mathbf{M} = (M_x, M_y, M_z)$ and its equilibrium state vector is $\mathbf{M}_{\mathrm{0}} = (0,0,M_0)$, where $M_0$ is the equilibrium spin magnetization along the $z$ axis. The dynamics follows the Bloch equations
\begin{equation}
	\frac{d\mathbf{M}}{dt} = \gamma \mathbf{M} \times \mathbf{B} - \mathbf{R}(\mathbf{M} - \mathbf{M}_{\mathrm{0}}),
	\label{eq:bloch-vector}
\end{equation}
where $\gamma$ is the gyromagnetic ratio and $\mathbf{R} = \operatorname{diag}(1/T_2, 1/T_2, 1/T_1)$ is the relaxation matrix. 
Here $T_1$ and $T_2$ correspond to the longitudinal and transverse spin relaxation times, respectively.

In the artificial feedback spin maser, the feedback field is produced by the transverse spin magnetization itself, which means that the transverse magnetization $\mathbf{M}_\perp = (M_x, M_y)$ is detected, amplified by a factor of $\alpha_{\mathrm{fb}}$ and fed back after a time delay of $\tau$. In this case, the feedback field has also different configurations of polarity and the geometry. The polarity is set by the sign of $\alpha_{\mathrm{fb}}$: $\alpha_{\mathrm{fb}}>0$ means positive feedback, while $\alpha_{\mathrm{fb}}<0$ means negative feedback. The geometric configuration is given by a set of binary switches $\sigma_{qp}\in\{0,1\}$, where $q,p\in\{x,y\}$ label the feedback axis and detection axis, respectively. $\sigma_{qp}=1$ activates the configuration that detects $M_q$ and feeds back along the $p$ direction, whereas $\sigma_{qp}=0$ deactivates the corresponding channel. Therefore, the feedback field $\mathbf{B}^{\mathrm{fb}}$ can be described by
\begin{equation}
	\begin{aligned}
		\mathbf{B}^{\mathrm{fb}}(t) &=
		\alpha_{\mathrm{fb}}\,
		\begin{pmatrix}
			\sigma_{xx}\quad\sigma_{xy}\\
			\sigma_{yx}\quad\sigma_{yy}
			\end{pmatrix}
		\begin{pmatrix}
		M_x(t-\tau)\\
		M_y(t-\tau)
		\end{pmatrix}.
	\end{aligned}
	\label{eq:feedback}
\end{equation}

In this paper, two typical kinds of geometric configurations are considered: \textit{in-phase}, where both detection and feedback are along the $x$ direction, defined by $\sigma_{xx}=1$ with all other $\sigma_{qp}=0$, and \textit{quadrature}, where detection is along the $x$ direction and feedback is along the $y$ direction, defined by $\sigma_{xy}=1$ with all remaining $\sigma_{qp}=0$. Combined with the two possible signs of $\alpha_{\mathrm{fb}}$, these yield four distinct feedback configurations: in-phase positive feedback, in-phase negative feedback, quadrature positive feedback, and quadrature negative feedback. The other kinds of feedback configurations, for example, detection and feedback along the $y$ direction, can be analyzed in the same way.

To characterize the feedback strength and delay in a unified manner, we introduce the dimensionless feedback strength $\alpha \equiv \alpha_{\mathrm{fb}} M_0 / B_0$ and the phase delay \(\theta \equiv \omega_0 \tau\), where $\omega_0 = |\gamma B_0| = 2\pi\nu_0$ is the Larmor frequency and $\nu_0$ is the Larmor frequency in Hz.
The static bias field is set to $B_0 = 3.0 \times 10^{-6}\,\text{T}$. Numerical simulations are carried out for $^{129}$Xe nuclear spin, which has a gyromagnetic ratio of $\gamma = -2\pi \times 11.78\,\text{MHz/T}$~\cite{Ratcliffe1998} and gives a Larmor frequency of $\nu_0 \sim 35.3\,\text{Hz}$. Xe nuclear spin magnetization can reach a relaxation time of $\sim$ 5 seconds in a $100\,^\circ\text{C}$ vacuum glass cell.  All simulations start from the initial spin state \(\mathbf{M}_\text{init} = \left(10^{-4}M_0,\,0,\,M_0\right)\).

\section{Results and Analysis}

Using the feedback-driven Bloch equations~\eqref{eq:bloch-vector} and the feedback configurations~\eqref{eq:feedback}, we did numerical simulations of the spin dynamics over a wide range of the \((\alpha,\theta)\) parameter space and analyzed the spin dynamics according to the time domain spin oscillations and the Fourier transformed spectra. 

First, we characterize the emergent dynamical regimes using the maximum Lyapunov exponent $\lambda_{\max}$ of the simulated time domain spin oscillations. For \(\lambda_{\max}<0\), all Lyapunov exponents take negative values, and the system undergoes damped oscillation converging to a stable fixed point where the oscillation amplitude gradually vanishes. 
For \(\lambda_{\max}\sim0\) with all remaining exponents negative, periodic self-sustaining oscillations emerge, corresponding to a stable limit cycle characterized by a constant amplitude and a single fundamental frequency. 
When at least two exponents equal zero and all others stay negative, quasiperiodic oscillation arises. Superposition of multiple incommensurate frequencies generates non-closing bounded trajectories that never exactly repeat the previous one. At last, \(\lambda_{\max}>0\) indicates the presence of chaotic dynamics, distinguished by aperiodic, irregular trajectories and extreme sensitivity to infinitesimal initial perturbations.

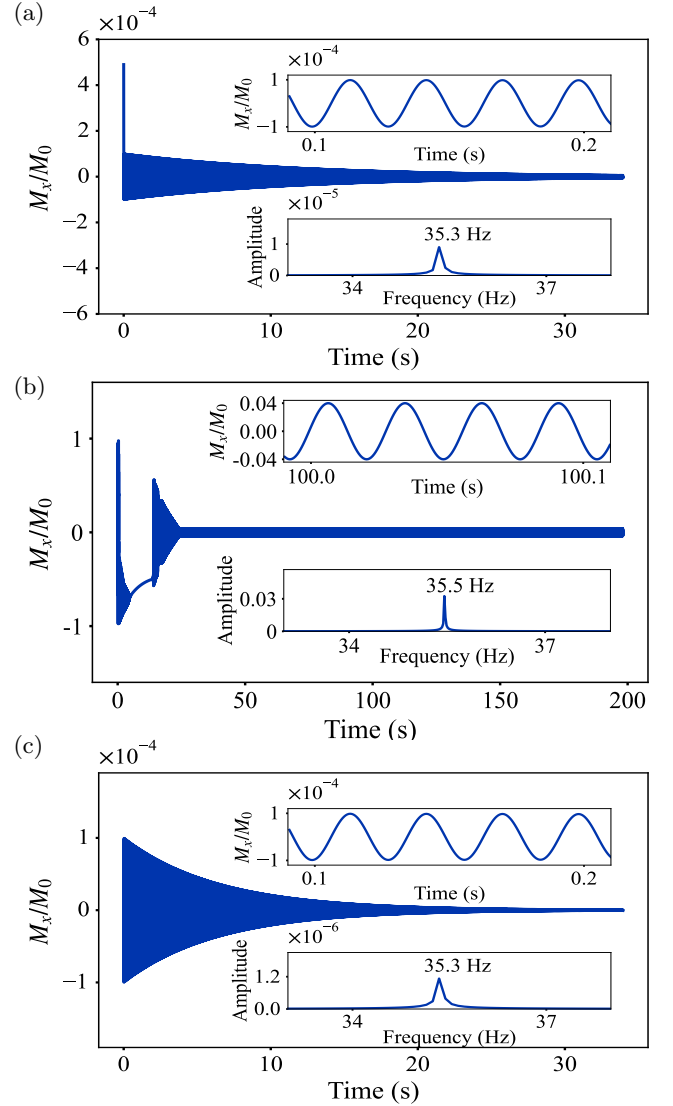
\begin{figure}[htbp]
	\centering
	\setlength{\unitlength}{\textwidth}
	
	\begin{picture}(0.5,0.27)
		\put(0.01,0){\includegraphics[width=0.46\textwidth,height=0.27\textwidth]{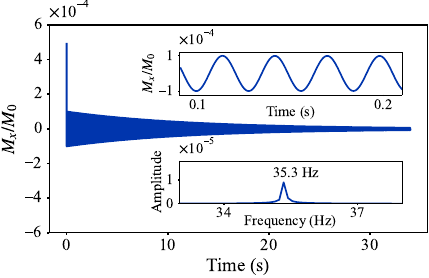}}
		\put(0.0,0.26){\small (a)\label{fig:Time-series_a}}
	\end{picture}\hfill
	
	\vspace{2pt}
	
	\begin{picture}(0.5,0.27)
		\put(0.01,0){\includegraphics[width=0.46\textwidth,height=0.27\textwidth]{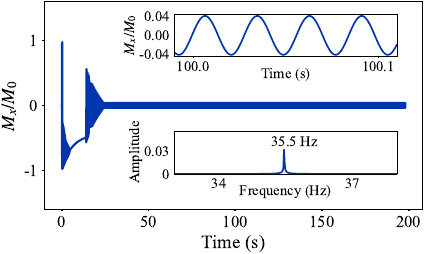}}
		\put(0.0,0.26){\small (b)\label{fig:Time-series_b}}
	\end{picture}\hfill
	
	\vspace{-3.3pt}
	
	\begin{picture}(0.5,0.27)
		\put(0.01,0){\includegraphics[width=0.46\textwidth,height=0.27\textwidth]{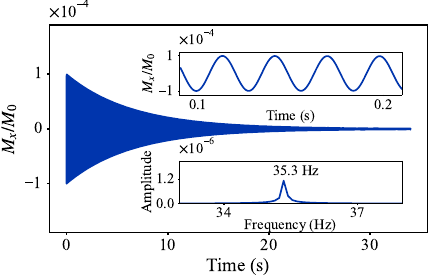}}
		\put(0.0,0.26){\small (c)\label{fig:Time-series_c}}
	\end{picture}\hfill
	
	\caption{Three typical oscillation modes of the artificial feedback spin masers: (a) In-phase positive feedback, weak feedback \(\alpha=0.001\) and \(\theta = 19\pi/18\), damped spin oscillation at Larmor frequency \(\nu_0\); (b) In-phase positive feedback, strong feedback \(\alpha=3.5\) and \(\theta = 19\pi/18\), self-sustaining oscillation near the Larmor frequency \(\nu_0\); (c) Quadrature negative feedback, weak feedback \(\alpha=-0.001\) and \(\theta = \pi/3\), damped spin oscillation at Larmor frequency \(\nu_0\).}
	\label{fig:Time-series}

\end{figure}

In the weak feedback regime (\(|\alpha|\ll 1\)),  all feedback configurations exhibit only two types of dynamical behavior---damped oscillation or periodic self-sustaining oscillation near the Larmor frequency \(\nu_0\), as shown in Fig. \ref{fig:Time-series}. In this case, the oscillation frequency shift \(|\nu-\nu_0|\) is much smaller than both \(\nu_0\) and the transverse relaxation rate \(1/T_2\), and the feedback delay time $\tau$ is much shorter than relaxation times $T_1$ and $T_2$; the Bloch equations can be solved by applying the rotating-wave approximation and the slowly-varying-amplitude assumption \cite{Wang2017,Feng2025}. From this treatment we can derive the threshold for making a successful spin maser, i.e., self-sustaining spin oscillations without obvious decay in the long-time operation. For in-phase feedback, the threshold depends on \(\sin\theta\). Specifically, for in-phase positive feedback (\(\alpha>0\)), the maser condition is
\begin{equation}
	\sin\theta < \frac{2}{\gamma B_0 T_2 \alpha}.
	\label{eq:xx_pos}
\end{equation}
As \(\gamma<0\), it needs \(\sin\theta<0\) for the above relation to hold; therefore the phase range for the spin maser is \(\theta\in(\pi,2\pi)\). The regime of the spin maser is symmetric about \(\theta=3\pi/2\) within this phase range. For in-phase negative feedback (\(\alpha<0\)), the maser condition is
\begin{equation}
	\sin\theta > \frac{2}{\gamma B_0 T_2 \alpha},
	\label{eq:xx_neg}
\end{equation}
which requires \(\sin\theta>0\); thus the phase range for the spin maser is \(\theta\in(0,\pi)\) and the spin maser regime is symmetric about \(\theta=\pi/2\) in this phase range.

For quadrature feedback, the intrinsic $\pi/2$ delay between the detection and feedback channels replaces \(\sin\theta\) by \(\cos\theta\) in the threshold formula. The maser condition for positive feedback is given by
\begin{equation}
	\cos\theta > -\frac{2}{\gamma B_0 T_2 \alpha},
	\label{eq:xy_pos}
\end{equation}
which, given \(\gamma<0\), leads to \(\cos\theta>0\). The oscillation intervals are \(\theta\in(0,\pi/2)\cup(3\pi/2,2\pi)\). The maser condition for negative feedback is
\begin{equation}
	\cos\theta < -\frac{2}{\gamma B_0 T_2 \alpha},
	\label{eq:xy_neg}
\end{equation}
which similarly yields \(\cos\theta<0\); the oscillation interval is \(\theta\in(\pi/2,3\pi/2)\), and the threshold curve is symmetric about \(\theta=\pi\) within this interval.

Under the condition of weak feedback, the phase delay \(\theta\) determines the oscillation threshold and the symmetry properties. 
For a given \(\theta\), the threshold condition yields the minimum \(|\alpha|\) required to excite self-sustaining oscillations. 
Reversing the feedback polarity (changing \(\alpha\) to \(-\alpha\)) shifts the oscillation phase interval by $\pi$; switching the geometric configuration from in-phase to quadrature introduces an intrinsic $\pi/2$ phase delay and transforms the threshold function from a sine to a cosine. These two operations correspond to translations in the phase dimension. 

\begin{figure}[htbp]
	\centering
	\setlength{\unitlength}{\textwidth}
	\begin{picture}(0.5,0.27)
		\put(0.01,0){\includegraphics[width=0.46\textwidth,height=0.27\textwidth]{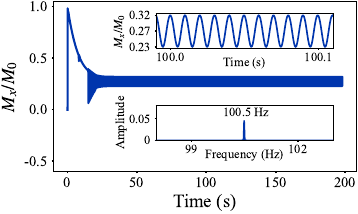}}
		\put(0.0,0.26){\small (a)\label{fig:New-frequency_a}}
	\end{picture}
	\begin{picture}(0.5,0.27)
		\put(0.01,0){\includegraphics[width=0.46\textwidth,height=0.27\textwidth]{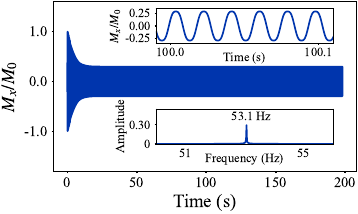}}
		\put(0.0,0.26){\small (b)\label{fig:New-frequency_b}}
	\end{picture}\hfill
\caption{New spin oscillation modes at specific feedback configurations: (a) in-phase positive feedback, strong feedback \(\alpha=9\) and \(\theta = 19\pi/18\), spin maser oscillates at 100.5 Hz; (b) quadrature negative feedback, strong feedback \(\alpha=-6\) and \(\theta = \pi/3\), spin maser oscillates at 53.1 Hz. Both frequencies are neither harmonics nor subharmonics of Larmor frequency.}
	\label{fig:New-frequency}
\end{figure}

In specific feedback configurations, we identify a novel stable oscillation mode, which exhibits the unique features of time crystals. Taking the in-phase positive feedback configuration at \(\theta=19\pi/18\) and the quadrature negative feedback configuration at \(\theta=\pi/3\) as examples, for $\left|\alpha\right|>1$, i.e. the feedback field stronger than the static field, a Hopf bifurcation generates a stable limit cycle. The system enters a self‑sustaining periodic oscillation at frequencies far away from the Larmor frequency $\nu_0$ and the new frequency is neither a harmonic nor a sub-harmonic of \(\nu_0\), as shown in Fig.~\ref{fig:New-frequency}.

By solving the steady state solution of Eq. \ref{eq:bloch-vector}, we identify two different kinds of stable limit cycles. The first kind of limit cycle converges to a nonzero transverse spin magnetization given by   
$M_x^* = \pm\sqrt{\frac{M_0 (M_0 - M_z^*)}{\alpha\, T_1 / T_2}}$, with $M_z^* = \frac{M_0}{\alpha}\left(1 + \frac{1}{\gamma^2 B_0^2 T_2^2}\right)$. This means the spin oscillations may have two different evolution paths determined by the initial state of $M_x$. The second kind of limit cycle converges to a zero transverse spin magnetization, meaning a single evolution path. Beyond distinct evolution paths, the two kinds of new limit cycles differ in convergence speed and steady-state amplitude. For the in-phase positive feedback, the system undergoes a complex transient before settling into the steady state, with a small steady-state amplitude. For the quadrature negative feedback, the system approaches the steady state more rapidly with a smoother transient response, and the steady-state amplitude is large. 

Despite these different dynamical features, both oscillation modes share the common features of time crystal: spontaneous symmetry breaking and robustness against perturbations. The spontaneous symmetry breaking is usually indicated by the random distribution of initial phases for spin oscillations in many time simulations. The robustness against perturbations is indicated by the independence of amplitude and frequency of spin oscillations on the random noise input for many times simulations. 

\begin{figure}[htbp]
	\centering
	\setlength{\unitlength}{\textwidth}
	\begin{picture}(0.5,0.6)
		\put(0,0){\includegraphics{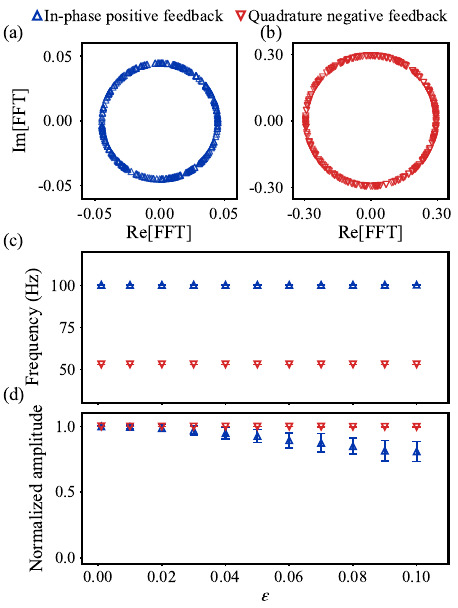}}
	\end{picture}
	\caption{(a) and (b) show the complex Fourier amplitudes of the new spin oscillations for the in-phase positive and quadrature negative feedback configurations in 300 independent simulations, respectively. These randomly distributed initial phases indicate the spontaneous symmetry breaking nature of the new spin oscillations. (c) and (d) show the frequency and amplitude of the new spin oscillations versus fluctuation in terms of a parameter $\varepsilon$. The simulation results show that these new spin oscillation modes exhibit strong robustness against longitudinal bias field fluctuations. The parameter $\varepsilon$ is swept from 0 to 0.1 in steps of 0.001. Each data point is obtained by averaging over 100 independent simulations, and the error bars represent the standard deviations. When the standard deviation tends to 0, the error bar is represented by a short horizontal line centered at the data point.}	
	\label{fig:robustness}
\end{figure}

For the former, we apply small perturbations $\delta_1$ and $\delta_2$ to the transverse magnetization at $t=0$ to mimic the unavoidable noise present in realistic experiments, such that $\mathbf{M}^{\delta}(0) = (\delta_1, \delta_2, M_0)$.
We perform 300 times of independent simulations. For each simulation, $\delta_1$ and $\delta_2$ are independently sampled from a zero-mean Gaussian distribution with standard deviation $\delta_0 = 10^{-7} M_0$.
Fig.\hyperref[fig:robustness]{~\ref*{fig:robustness}(a)} and Fig.\hyperref[fig:robustness]{~\ref*{fig:robustness}(b)} show that the complex Fourier amplitudes of all trajectories are distributed on a circle in the complex plane. This indicates that the macroscopic oscillation amplitude and frequency are preserved, while the phase is uniformly distributed over \([0,2\pi)\) across independent realizations. Such a uniform phase distribution is a signature of spontaneous symmetry breaking.

For the latter, we apply a randomly fluctuating magnetic field $\delta B_z$ into the static bias field $B_0$, such that the overall longitudinal field $\tilde{B}_z = B_0 + \delta B_z$. At every timestep in the numerical simulations, $\delta B_z$ is randomly sampled over the range $[-\varepsilon B_0, \varepsilon B_0]$, where the dimensionless parameter $\varepsilon$ denotes the fluctuation strength.
Fig.\hyperref[fig:robustness]{~\ref*{fig:robustness}(c)} shows that the peak frequency of these new spin oscillation modes stays at their unperturbed values across the interval $\varepsilon \in [0, 0.1]$. Fig.\hyperref[fig:robustness]{~\ref*{fig:robustness}(d)} shows that the normalized amplitude of these new spin oscillation modes at specific feedback configurations drops slowly with $\varepsilon$. For the in-phase positive feedback configuration, the amplitude of new spin oscillations goes down to 0.7 at maximum fluctuation $\varepsilon=0.1$. In contrast, for the quadrature negative feedback configuration, the amplitude remains nearly invariant throughout the interval $\varepsilon \in [0, 0.1]$. In general, both new spin oscillations exhibit strong robustness against the perturbations in the longitudinal bias field.
The above simulation results prove that both oscillation modes satisfy the criteria for time crystals~\cite{Wilczek2012, Keßler2021, YiThomas2024}.

\begin{figure*}[htbp]
	\centering
	\includegraphics[width=\textwidth]{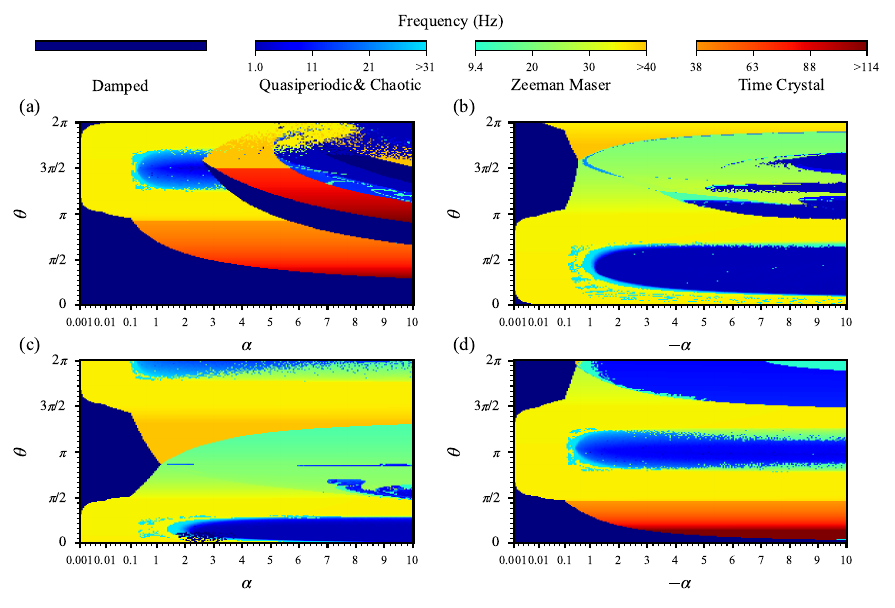}
	\caption{Dynamical phase diagrams for the four kinds of feedback configurations: (a) in-phase positive feedback---$x$-detection, $x$-feedback, $\alpha>0$; (b) in-phase negative feedback---$x$-detection, $x$-feedback, $\alpha<0$; (c) quadrature positive feedback---$x$-detection, $y$-feedback, $\alpha>0$; (d) quadrature negative feedback---$x$-detection, $y$-feedback, $\alpha<0$. While normal nonlinear spin dynamics including the quasiperiodic and chaotic oscillations exist for all four kinds of feedback configurations, the time crystal effects present only at the first and last kinds of feedback configurations. These simulations results agree with our former simulations \cite{Feng2025} and experiments on time crystals \cite{Wang2025}. Note: The horizontal axis of feedback strength $\alpha$ uses a piecewise linear scale---for $0.001 \le \alpha \le 1$, major ticks are placed at successive powers of ten ($0.001$, $0.01$, $0.1$, $1$), while equally spaced major ticks are used for $1 < \alpha \le 10$.}
	\label{fig:phase_diagram}
\end{figure*}

To summarize, we performed full parameter space simulations for the four typical kinds of feedback configurations, as shown in Fig.~\ref{fig:phase_diagram}. For weak feedback \(|\alpha|<1\), the spin dynamics is relatively simple, falling into two kinds of spin oscillations: damped oscillations and spin maser oscillations. For strong feedback \(|\alpha|>1\), the system enters the nonlinear regime with rich nonlinear dynamics. For different feedback configurations, the spin dynamics depend not only on strength \(\alpha\) and the phase delay \(\theta\) of the feedback field, but also on the geometry configuration, i.e. the polarity and detection-feedback geometry, which was not paid much attention to in earlier studies. 

The combined effect of feedback polarity and geometry cannot be described by simple linear superposition, and each combination yields unique dynamical regulation. Under in-phase positive feedback, the system exhibits the most complicated evolution routes: the spin dynamics can cross many times between damped oscillation, Zeeman maser, time crystal, quasiperiodic oscillation and chaos as the feedback strength and phase delay change. For the in-phase negative feedback configuration, the system follows a simpler spin dynamical evolutionary route and agree with our previous simulation results \cite{Feng2025}. The quadrature positive feedback configuration favors most the realization of Zeeman maser and limits the occurrence of nonlinear dynamics including the quasiperiodic oscillation and chaos. For the quadrature negative feedback configuration, the system has simpler spin dynamics than it does for the in-phase positive feedback, but the time crystal effects are favored in terms of the signal amplitude and robustness against perturbations as discussed before.

Specifically, the time crystal effects appear only in in-phase positive and quadrature negative feedback configurations, indicating the importance of  feedback configurations. Within the time crystal region, the oscillation frequency remains nearly invariant as the feedback strength \(\alpha\) increases. In contrast, as the phase delay \(\theta\) increases, the oscillation frequency decreases, yielding a linear growth of the oscillation period \(T\) with \(\theta\) (where \(T=1/f\), note the frequency color bar for time crystal). This distinct parameter dependence demonstrates a clear separation of roles between the two control parameters: \(\alpha\) dominates the dynamical phase transition behavior and tunes the oscillation amplitude, whereas \(\theta\) determines the intrinsic oscillation frequency of the time crystal. These results also agree with our recent experimental observations~\cite{Wang2025}.

\section{Conclusion}

In conclusion, we investigated theoretically the dependence of nonlinear dynamics on the feedback configurations in an artificial feedback spin maser. By tuning the detection-feedback geometry, strength and phase delay of the feedback field, we show that the spin maser has different nonlinear dynamics including quasiperiodic and chaotic spin oscillations other than the intrinsic Larmor precession. For the four typical kinds of feedback configurations, we presented the corresponding spin maser criteria and full range phase diagrams, which are very useful in designing and operating the artificial spin masers in practice. In this context, new modes of spin oscillations were found at two specific configurations: in-phase positive feedback and quadrature negative feedback. The new spin oscillations exhibit features consistent with the definition of time crystals: emergent non-Larmor oscillation frequency, strong robustness against noise, and spontaneous phase randomization under minute perturbations. This work presents a different viewpoint on the recently observed ``time crystal'' effects in artificial feedback spin masers \cite{Tang2024arxiv,Wang2024CP,Wang2025,Huang2025NC}.

\bibliography{reference}
	
\end{document}